\documentclass[12pt]{article}  
  \usepackage{psfig,times,fancyheadings}
        
\begin{document}
\thispagestyle{empty}

  \lhead[\fancyplain{}{\sl }]{\fancyplain{}{\sl }}
  \rhead[\fancyplain{}{\sl }]{\fancyplain{}{\sl }}

%%%%% Definitions

\newcommand{\nc}{\newcommand}

\nc{\qI}[1]{\section{ {#1} }}
\nc{\qA}[1]{\subsection{ {#1} }}
\nc{\qun}[1]{\subsubsection{ {#1} }}
\nc{\qa}[1]{\paragraph{ {#1} }}

\nc{\qfoot}[1]{\footnote{ {#1} }}
\def\qL{\hfill \break}
\def\qpar{\vskip 2mm plus 0.2mm minus 0.2mm}

\def\qtvi{\vrule height 2pt depth 5pt width 0pt}
\def\qth{\vrule height 12pt depth 0pt width 0pt}
\def\qtb{\vrule height 0pt depth 5pt width 0pt}
\def\tvi{\vrule height 12pt depth 5pt width 0pt}

\def\qci#1{\parindent=0mm \par \small \parshape=1 1cm 15cm  #1 \par
                \normalsize}

\def\qparr{ \vskip 1.0mm plus 0.2mm minus 0.2mm \hangindent=10mm
\hangafter=1}

             % Enumerations
\def\qbu{\hfill \par \hskip 6mm $ \bullet $ \hskip 2mm}
\def\qee#1{\hfill \par hskip 6mm #1) \hskip 2 mm}

                 % Decale UN paragraphe
                 % Attention! La double accolade est vitale, sinon tout le
                 % est decale (cf TEX p.199)
                 % On peut aller a la ligne avec \qL=\hfill \break
                 % Par contre ne supporte pas les lignes blanches
\def\qdec#1{\par {\leftskip=2cm {#1} \par}}

    %% Defs specifiques
\def\qdpt{\partial_t}
\def\qdpx{\partial_x}
\def\qddpt{\partial^{2}_{t^2}}
\def\qddpx{\partial^{2}_{x^2}}
\def\qn#1{\eqno \hbox{(#1)}}
\def\qds{\displaystyle}
\def\qal{\sqrt{1+\alpha ^2}}
\def\qw{\widetilde}

            %% Legendes figures
\def\qhu{\hskip 0.6mm}
\def\qhv{\hskip 3mm}
\def\qhw{\hskip 1.5mm}
\def\qv{\vskip 0.1mm plus 0.05mm minus 0.05mm}

\def\qleg#1#2#3{ \noindent
{\bf \small #1\qhw}{\small #2\qhw}{\it \small #3}\qv }

%%%%% End of definitions

\null
\vskip 1cm

\centerline{\bf \LARGE Response Functions
to Critical Shocks in Social Sciences:}
\centerline{\bf \LARGE An Empirical and Numerical Study}
\vskip 0.5cm

\vskip 1cm
\centerline{\bf B. M. Roehner$^1$, D. Sornette$^{2,3}$ and J.V. 
Andersen$^{3,4}$}

\vskip 1cm

PACS: 01.75+m Science and society - 05.40+j Fluctuations phenomena -
05.70.Jk Critical phenomena
\qL
Keywords: Response functions; Critical shock; Cohesion; Collective
behavior; Observational strategy; Quasi-experimental.

\vskip 1cm

{\bf Abstract}\quad 
We show that, provided one focuses on properly
selected episodes, one can apply to the social sciences
the same observational strategy that has proved successful in
natural sciences such as astrophysics or geodynamics.
For instance, in order to probe the cohesion of a policy, one
can, in different countries, study
the reactions
to some huge and sudden exogenous shocks, which we call Dirac shocks.
This approach naturally leads to the notion of structural (as
opposed or complementary to temporal) forecast.
Although structural predictions are
by far the most common way to test theories in the natural sciences,
they have been much less used in the social sciences. The
Dirac shock approach opens the way to testing structural
predictions in the social sciences. The examples reported here
suggest that critical events are able to reveal
pre-existing ``cracks'' because they probe the social
cohesion which is an indicator and predictor of future evolution of the system,
and in some cases foreshadows a bifurcation. We complement
our empirical work with numerical simulations of the response function
(``damage spreading'') to Dirac shocks in
the Sznajd model of consensus build-up. We quantify the
slow relaxation of the difference between perturbed
and unperturbed systems, the conditions under which the consensus
is modified by the shock and the large variability from one
realization to another.

\vskip 1cm

1: Institute for Theoretical and High Energy Physics
\qL
\phantom{1: }Postal address: LPTHE, University Paris 7, 2 place Jussieu,
75005 Paris, France.
\qL
\phantom{1: }E-mail: roehner@lpthe.jussieu.fr
\qL
\phantom{1: }FAX: 33 1 44 27 79 90
\qpar

2: IGPP and ESS Department, University of California, Los Angeles,
California 90095.

\qpar
3: LPMC, CNRS UMR6622 and Universit\'e des Sciences, BP 70, Parc Valrose,
06108 Nice Cedex 2, France.
\qL
\phantom{3: }E-mail: sornette@unice.fr

\qpar
4: U. F. R. de Sciences \'Economiques, Gestion, Math\'ematiques et
Informatique, \\ CNRS UMR7536 and Universit\'e Paris X-Nanterre,
92001 Nanterre Cedex, France
\qL
\phantom{4: }E-mail: vitting@unice.fr

\vfill \eject

\qI{Posing the problem}

Self-organized criticality, and more generally, complex system theory
contend that out-of-equilibrium slowly driven systems with threshold dynamics
relax through a hierarchy of avalanches of all sizes. Accordingly,
extreme events are seen to be endogenous, in contrast with previous
prevailing views. But, how can one assert with 100\% confidence that a given
extreme event is really due to an endogenous self-organization of the
system, rather than to the response to an external shock? Most natural
and social systems are indeed continuously subjected to external stimulations,
noises, shocks, sollications, forcing, which can widely vary
in amplitude. It is thus not clear a priori if a given large event is
due to a strong exogenous shock, to the internal dynamics of the
system, or maybe to a combination of both. Adressing this question
is fundamental for understanding the relative importance of
self-organization versus external forcing in complex systems.
This question, whether distinguishing properties
characterize endogenous versus exogenous shocks, permeates
many systems, for instance,  biological extinctions such as the
Cretaceous/Tertiary KT boundary (meteorite versus extreme volcanic
activity versus self-organized critical extinction cascades),
commercial successes
(progressive reputation cascade versus the result of a well
orchestrated advertisement),
immune system deficiencies (external viral/bacterial infections
versus internal cascades of regulatory breakdowns), the aviation
industry recession (9/11 versus structural endogenous problems),
discoveries (serendipity versus the outcome of slow endogenous
maturation processes),
cognition and brain learning processes (role of
external inputs versus internal self-organization and
reinforcements for instance during dreams) and wars
(internally generated
(civil wars) versus imported from the outside) and so on.
In economics, endogeneity versus exogeneity has been
hotly debated for decades. A prominent example is the
theory of Schumpeter on the importance of technological
discontinuities in economic history (Schumpeter, 1939).
Schumpeter argued that ``evolution is
lopsided, discontinuous, disharmonious by nature... studded with
violent outbursts and catastrophes... more like a series of
explosions than a gentle, though incessant, transformation''.
Thus, by emphasizing the endogenous nature of jumps in economic evolution,
Schumpeter de-emphasized the new for exogenous shocks to
explain major disruptions.
Endogeneity versus exogeneity is also
paramount in economic growth theory (Romer, 1996).
\qpar

In general, the time evolution of an open out-of-equilibrium social system
can often be seen as the result of its intrinsic nonlinear dynamics
subjected to a forcing which can be schematically divided
into two classes: (1) a flux of small noise-like fluctuations
and (2) jumps or Dirac impulses strongly perturbing the system
over a time scale much shorter than its characteristic internal time scales.
In the former case (1), the system may exhibit spontaneously
different regimes, with burst of activity not directly related
to the intensity of the forcing. This is refered to as an
``endogenous'' activity. In the later case (2), the Dirac impulse
gives rise to a reaction of the system which would correspond to
the Green function or propagator response function if the system was
linear, to use
the terminology of physics. This response function corresponds
to a stimulus whose origin is clearly exogenous.
Beyond these two end-member cases (1) and (2), most of the time,
endogenous and exogenous effects
are intimately entangled. Again, if the system dynamics is
linear, it is possible to predict the endogenous dynamics from the
calibration of the exogenous response function. This idea
has been pursued in (Sornette et al., 2003; Johansen and Sornette, 2004)
for applications to financial stock markets, in (Sornette et al., 2004)
for applications to commercial sales and in (Helmstetter et al., 2003)
for applications to earthquakes (see
(Sornette and Helmstetter, 2003 for a simple theory and numerical simulations
in the case of linear dynamics with long-range memory Green functions).
In the financial case, the Dirac impulses used in the analysis were the
9/11/2001 event and the coup against Gorbachev (Sornette et al., 2003)
and several other major disruptive political events such as wars
(Johansen and Sornette, 2004). In the case of commercial sales, the Dirac
impulses were very strong advertisements of books occurring during
major TV shows or printed in major newspapers (NYT).

Such exogenous Dirac impulse are rare. However,
once we have recognized and accepted the importance of
exogenous shocks in the dynamics of social systems,
we are in a much better position to use these shocks for improving
our understanding.
Instead of merely monitoring the system in a continuous way in the
course of time, we can
concentrate our attention on specific episodes during which the system
experienced such Dirac-type shocks.
During such episodes, the behavior of the system will be completely
dominated by its Green's function
response (to use a linear terminology). Momentarily, we can
ignore its stochastic features and background noise. This represents a
crucial simplification. Once the response of the system to a Dirac
shock has been calibrated, we are in a much better position to understand
the dynamics resulting from the interwoven flux of external
perturbations and internal organization. This results from the fact
that, in a linear framework, the dynamics of a system under arbitrary
conditions can be reconstructed by a convolution of the Green function
with the flux of externally imposed sollicitations (Morse and Feshbach, 1953).
A few recent papers show the value of also studying the response to
short impulses
in nonlinear dynamical systems, with applications to classical mechanics
(Dellago and Mukamel, 2003) and to macroeconomics (Potter, 2000).
\qpar

Considering therefore the importance of characterizing the response of
the system
to Dirac impulses, our purpose here is to focus on documenting
important historical cases of Dirac impulses. We do not intend to make
a detailed study of specific cases, however; rather, our objective is
to demonstrate the feasibility of the approach by emphasizing the
robustness of responses to Dirac impulses and by describing
appropriate sources.
The paper is organized as follows.
Section \ref{defshock} explains more precisely what we mean by a Dirac shock
by way of examples. In section \ref{nexle}, we show that, provided it
is duly exploited, the new possibility of following the development of
major social events in real time globally over the world marks
the beginning of a new
era in social research. Readers who are already convinced
of the importance of this progress may wish to skip this
epistemological discussion and move directly to the analysis
of actual examples in the following sections. Section \ref{exsoc}
makes an analogy between social cohesion and material cohesion, to
suggest possible ways of quantifying the former. Section
\ref{critical} has the objective of describing available observation
devices; in particular, we emphasize their new capabilities as
well as their limitations and what can
be learned about cracks between various communities.
Section \ref{criticalecon} then presents
examples of the method of critical events in economics.
Section \ref{criticalhisto} explains how
the method of critical events can be extended to pre-Internet times.
The last section concludes by
comparing the nature of structural versus temporal predictions.

\qI{Dirac shocks \label{defshock}}

In the early hours of December, 6, 1992, thousands of Hindus converged
toward the holy city of Ayodhya in northern India and began
to destroy the Babri mosque which was said to be built on the
birthplace of Lord Rama. The old brick walls came down fairly
easily and soon the three domes of the mosque crashed to the
ground. This event triggered a burst of protestations and
retaliations which swept the whole world from Bangladesh to
Pakistan, to England or the Netherlands. In all these
countries, Hindu people
were assaulted, Hindu temples were firebombed, damaged or destroyed.
%A parallel with a wave propagation phenomenon (which will be discussed in
%more detail below) immediately comes to mind:
%a stone is tossed into a pond, the corresponding splash forms
%a large wave around the entry point which then travels
%toward the far end of the pond and disturbs
%the reeds on the side of the pond.
\qpar

Throughout history there have been many episodes of that kind.
For instance, it is said that after hearing of the storming
of the Bastille that occurred on July, 14, 1789, the German philosopher
Immanuel Kant postponed his immutable afternoon walk. In those
days, however it was very difficult to observe the aftershocks
of big events in any detail.
There were no news agencies and only few newspapers.
Even in more recent times, although not strictly impossible,
such investigations were very difficult to carry out.
For instance, at the time of the assassination of president
John F. Kennedy (22 November 1963), news agencies
and newspapers were many, but assessing its impact by collecting and reading
them would have been a very demanding task if only because it
would have
required a knowledge of many languages that few social scientists
would have.
\qL
By 1992, the situation has radically
changed to the point that we are able to scrutinize the propagation of
the waves of reactions to a sudden shock swelling worldwide
and day by day. This opens a completely
new perspective in the social sciences; in the future, it will permit
quasi-experimental research and will bring social sciences much
closer to other observational sciences such as geophysics or
astrophysics. Possible implications are discussed in more
detail in this paper, but first of all, let us examine more
closely the factors which brought about such a methodological revolution.
As mentioned in the title, the Internet has been of course a crucial
factor, but it was not the only one. At the time of writing
(2003), softwares for the automatic translation of
languages are still very crude to the point of making
translations unclear and obscure. As a result, the language
barrier would be as serious now as it was in 1963 if there had
not be another change, namely the fact that English has
become a lingua franca, that is to say a language of communication
that is used worldwide; as a result, all major news agencies,
whether they are Chinese, French, German, Indian, Japanese
or Russian deliver their dispatches and news wires simultaneously
in their national language and in English. Thanks to this circumstance,
we are now able to observe events
that occur in different parts of the world with fairly good precision.
In summary, for the first time in history, we can now watch
major collective movements across the globe, day by day
or even hour by hour. This is used or over-used by some television
channels which now make a business at exploiting this flux of news
to captive and capture their audience. This has led to
a surge of journalists killed due to the growing media coverage
in extremely dangerous conditions (cf the 2003 Iraq war).

\qI{Why the quasi-experimental approach provides a more efficient
perspective \label{nexle}}

It is often claimed that natural phenomena are simpler than
social phenomena and that it is this complexity gap which accounts
for the fact that the social sciences have been less successful.
True, it is awfully difficult to describe social systems {\it in
detail} but it is almost as difficult to describe a beam
of wood (say of oak) {\it in detail} from quark up to molecular level.
However, this did not prevent
physicists from {\it measuring} the speed of sound in a beam of wood
and to compare it to the speed of sound in other materials such
as steel, water or air.
This, in a nutshell, is the argument that we develop in this
section and in the next. But first of all, let us recall some
of the main characteristics of the different approaches that have
been used in the social sciences; this will help us realize
that the approach proposed in this paper is a natural
extension of the comparative perspective already used by
some social scientists.
\qpar

Traditionally, there have been two main perspectives in
the social sciences, the descriptive approach and the
problem-oriented approach
\qfoot{A French social scientist, Professor Pierre
Rosanvallon, refers to these two approaches as
``history as
remembrance'' ({\it l'histoire-m\'emoire}) and ``history
as a laboratory'' ({\it l'histoire-laboratoire}).}.
The first approach consists in describing a specific
social system with as much accuracy and detail as
permitted by available sources. This approach is still
dominant nowadays in many fields, for example in history,
economic history, or even sociology. As an illustration,
consider two recent studies bearing the following titles:
{\it Culture and inflation in Weimar Germany} or
{\it The first century of United States steel
corporation 1901-2001} (these are two recently
published books chosen almost at random in the
field of economic history). Needless to say, such one-period,
one-country studies cannot have any comparative claim or
ambition.

\qL
As an illustration of the problem-oriented approach,
one can mention Durkheim's celebrated study of
suicide (Durkheim 1897). What made
Durkheim's comparative approach possible (and this is
really the crux of the matter) is the fact that a suicide
in England and a suicide in Prussia can  be
considered as two manifestations of the {\it same
phenomenon}. Without that assumption, it would not
make sense to use data from both places in a comparative
way. Because suicide is a very basic phenomenon, nobody
would seriously object to that
hypothesis of universality. However, for other social phenomena,
universality may be less obvious. For instance,
are the stock market crashes of 1881 in Paris and
of October 1929 in New York two manifestations
of the same phenomenon. The present authors
(see for instance Roehner 2002, Sornette 2002 and
references therein) are among those who
find the universality assumption compelling while others
may consider that these events are unique and can not
therefore be compared in any meaningful way.
In short, if the descriptive approach is taken,
we may not be able to apply
the problem-oriented approach.

\qL
One must admit that the boom and bust of a
stock market is not as clearly defined as a suicide,
and the question of whether it makes sense to study
such events from a comparative perspective cannot be
settled by theoretical arguments. It is only by
showing that a new understanding can be gained from
the comparison that one will be able to justify it.
Naturally, the less basic a given phenomenon, the less
useful the comparative approach will be. If, instead
of focusing on the stock market crash of 1929, one wants
to include the subsequent economic crisis as well, then it is
less likely than a comparative approach will be successful
as more dimensions of the problems will make the comparison
less robust.
Contrariwise, the more basic the phenomenon under
consideration, the more useful the comparative approach.
The recently published books of the authors tried to
apply that philosophy to various historical events
from separatist disturbances to general strikes, to
wars of conquest (Roehner 2002 a,b) to the stock market (Sornette, 2002).
In each instance, the prerequisite is to define the
event under investigation as sharply as possible. Thus
for instance, because there are many kinds of general strikes,
it would be meaningless to study them comparatively;
however, it turns out that, for a particular kind
of general strikes that we call mushroom strikes,
there is a ``basic'' common underlying mechanism.
\qpar

Exactly the same problem arises when we try to implement
the quasi-experimental approach outlined at the beginning of
the paper. Indeed, it would be
of little interest to study the Ayodhya episode taken
alone; one wants to compare that episode with similar
ones in order to see if there is a pattern. Two examples
may be helpful at this point. (i) On October, 31, 1984, the
Prime minister of India, Indira Gandhi, was assassinated
by two of her Sikh bodyguards. This event triggered a
wave of retaliations against Sikh people and Sikh property, not only
in India (particularly in New Delhi), but in many other
countries as well. (ii) On September, 11, 2001, two planes
crashed into the twin towers of the World Trade Center in
New York. This event triggered a wave of reactions
against Islamic people and property
not only in the United States but also in other countries
(see below). Are these two phenomena of the same kind
as the one that occurred in the aftermath of the destruction
of the Babri mosque or is each event unique to the point
that comparing them would make little sense? Needless to say,
if one sides with the second conclusion, little benefit
can be gained from a comparative study; in that case the
resources of the Internet can be used only for the purpose
of building a catalog of individual events from which
little scientific knowledge will obtain. Now,
if we accept the assumption that it is worthwhile to
compare these events, a lot remains to be done. In
particular, as in physics,
we are faced with the difficult
task of addressing the ``right'' questions.
For instance, following the stone-in-the-pond allegory, one could try to
measure the vibrations of the reeds resulting from
the splash, but it is not obvious that this is the best
question to be asked. It would be more to the point
to measure the changes in water level if only
because the water is the same in all ponds, whereas
the physical characteristics of the
reeds may differ from one pond
to another.

\qL
In the examples that follow, we present some
quantitative evidence in order to convince the
reader of the potential of this approach (and also
in order to point out its limitations) but we
do not claim that the questions on which we focus
are the ``right''  ones. This approach is quite new
and it will probably take some time until its full
dividends can be reaped.

\qI{An example: how to measure social cohesion \label{exsoc}}

Recently, the definition and measure of social
cohesion has attracted a lot of attention. For instance,
in a much acclaimed study, Robert Putnam (2000) investigated
how social capital (see below) has changed in
the United States during recent decades, and Ashutosh
Varshney (2002) investigated the influence of social capital
on the antagonism between Hindus and Muslims in India.
Needless to say, social cohesion has many facets,
such as interactions at the family level, in the workplace, in
institutional groups (clubs, unions, political
parties or religious congregations: it is to this
aspect that the notion of social capital refers). Social cohesion
is also influenced by indirect interaction
brought about by being exposed to the same kind of
exogenous shocks (news, foreign threats, etc.). In
fact, the list of the factors which may contribute to
social cohesion is probably boundless. Yet, it is
important to realize that the cohesion of a piece of
wood, of steel or of a drop of water has many facets as
well. Writing the equation of state
even for a fairly simple system such as a
non-perfect gas turned out to be a very tricky problem.
%in his Nobel lecture (December, 12, 1910), Johannes Van
%der Waals goes as far as saying that ``an equation of
%state compatible with experimental data is totally
%impossible. No such equation is possible unless something
%is added, namely that the molecules associate to form larger complexes.''
This complexity, however, did not
prevent physicists from performing various measurements.
For instance, they measured the heat capacity or the
speed of sound for a vast set of substances ranging
from air to diamond. In return, these measurements
provided indirect estimates for internal parameters
of these substances. For the purpose of illustration,
let us consider the speed of sound; from theoretical
considerations, we know that the speed of sound in
a uniform medium is given by:
\begin{equation}
\hbox{Speed of sound [m/s]} =
\sqrt{{ \hbox{Bulk modulus of elasticity [N/m}^2\hbox{]}
\over \hbox{Density [kg/m}^3\hbox{]} }} %\qn{1} 
\end{equation}
\qpar

Once we have measured the speed of sound and once
the density is known from an independent experiment, this
formula can be used to define and measure the
bulk modulus of elasticity which can be considered
as a possible definition of the cohesion or hardness
of a medium. The figures given in the following table
confirm that this definition is indeed consistent with our
intuitive understanding of the cohesion of a substance
(the data refer to materials at room temperature and atmospheric
pressure).

$$ \matrix{
\tvi
  & \hbox{Hydrogen} & \hbox{Water}  & \hbox{Zinc} & \hbox{Iron} &
\hbox{Diamond} \cr
\noalign{\hrule}
\qth
\hbox{Bulk modulus of elasticity} \ [\hbox{N/m}^2]
  \hfill & 0.00015 & 2.2 & 72 &160 & 542 \cr
\hbox{Distance to nearest neighbor} \ [\hbox{Angstr\"om}]
  \hfill & 3000 & \sim 5 & & 2.5 & 1.5 \cr
\hbox{Velocity of sound} \ [\hbox{m/s}] \hfill & 1,300 & 1,500 & 4,200
& 6,000 & 12,000 \cr
} $$
Source: De Podesta (2002)
\vskip 3mm

In short, we may define the cohesion of a substance in the
following way:
$$ \hbox{Cohesion of a substance} = (\hbox{Speed of sound})^2
(\hbox{Density})$$

The same kind of reasoning can be used for social systems
which means that instead of describing all agencies which
contribute to social cohesion in each specific society, we
will measure some kind of diffusion speed and derive the
cohesion from this velocity.
\qpar

For the events that we have considered so far, the news
is spread by the media in a matter of minutes to hours. In other words
the interaction is not propagated between nearest neighbors;
it is more akin to the way an hormone is released in the
blood circulation and is global in nature. At this point,
we have the choice between
two different strategies. (i) We may be able to find shocks
which are transmitted through a neighbor-to-neighbor
interaction; commodity (spot) price shocks, for instance, are of that
kind (see Roehner 1999). In a more general way, what we
need are social phenomena which are not triggered merely
by information but by some kind of material effect.
(ii) If we insist on using information events (as we
do in this paper) then, we would need to replace
formula (1) by an expression suited to the presence
of non-local interactions, such as with the hormone-like effect.
\qpar

\qI{Response to  critical events \label{critical}}

In which sense do we use the expression ``critical event''?
An analogy may be useful. Suppose one wants to know whether
there is a crack in a girder of steel. A possible method
is to strike the girder with a hammer and to record the
vibrations. The crack will reveal itself in the way
the vibration pattern differs from the standard vibration
pattern of a flawless girder (this is an example
of the so-called non-destructive technique). A similar technique is used
in prospecting for oilfields. A shock is generated (usually
by using an explosive) at location $ A_0 $ and the resulting
vibrations are recorded at neighboring locations $ A_1, A_2, \ldots $.
By comparing these records to the vibration patterns
which would be expected without the presence of an oilfield
one might be able to derive its location, depth and size.
Similarly, we use here critical events as a mean for
locating cracks in social cohesion. The destruction of the
Babri mosque may be used as a hammer blow in order to reveal
possible cracks in the social fabric which
links Muslim and Hindu communities. In the same way, the
storming of the Golden Temple of the Sikhs (June 7, 1984) and
the subsequent assassination of Indira Gandhi may be used to
probe the relations between the Sikh and Hindu communities.
\qL

However, in this methodological paper, we will not go into
the details of these investigations. This section has the more
limited objective of describing available observation
devices; in particular, we emphasize their new capabilities as
well as their limitations. At the end we briefly discuss what can
be learned about cracks between various communities.
\qpar

The main device that we use is a database of newspaper
articles called Lexis-Nexis. Although it is not freely
available on the Internet, it is available in many
departments of political science or sociology. In the
Paris area (France), it is for instance available at the
National Foundation for Political Science (rue Saint
Guillaume). The data base goes back to the late
1970s and, not surprisingly, the yearly number of newspapers
available in the base increases in the course of time
along with the computerization of the newspapers industry. The base
can be searched by first defining a time window (for
instance 16 to 17 September 2001) and then by entering
a number of keywords which are to be found in the
articles. As an example, for the investigation
regarding the retaliations against Hindu temples in the
wake of the destruction of the Babri mosque, we used the
following combination of keywords:
\qpar

(temple OR hindu center OR hindu centre) AND (firebomb
OR firebombed OR torched OR arson attack OR ablaze OR
afire OR set on fire OR destroyed)

\qpar

For the aftermath of September 11, we replaced the keywords
within the first parenthesis by:
\qL
mosque OR mosques OR islamic school OR islamic center
OR islamic centre
\qpar

Two-day intervals were used to cover the 10 days before
the attack and the 20 days after the attack. Ten-day
intervals were used to cover the more distant dates.
In a first phase, we merely counted the number of articles
in each time interval without trying to read them. Used in
this way, Lexis-Nexis is in a sense similar to a particle
detector which merely records the number of particle impacts
without further documenting their characteristics, for instance
their energy and velocity.
\qL
The results are summarized in Fig.1 (thin and thick solid lines).

%%-----------------------------------------------
%%%% Fig.1
   \begin{figure}[htb]
     %\centerline{\psfig{width=17cm,figure=roehnersa1.eps}}
      \centerline{\psfig{width=17cm,figure=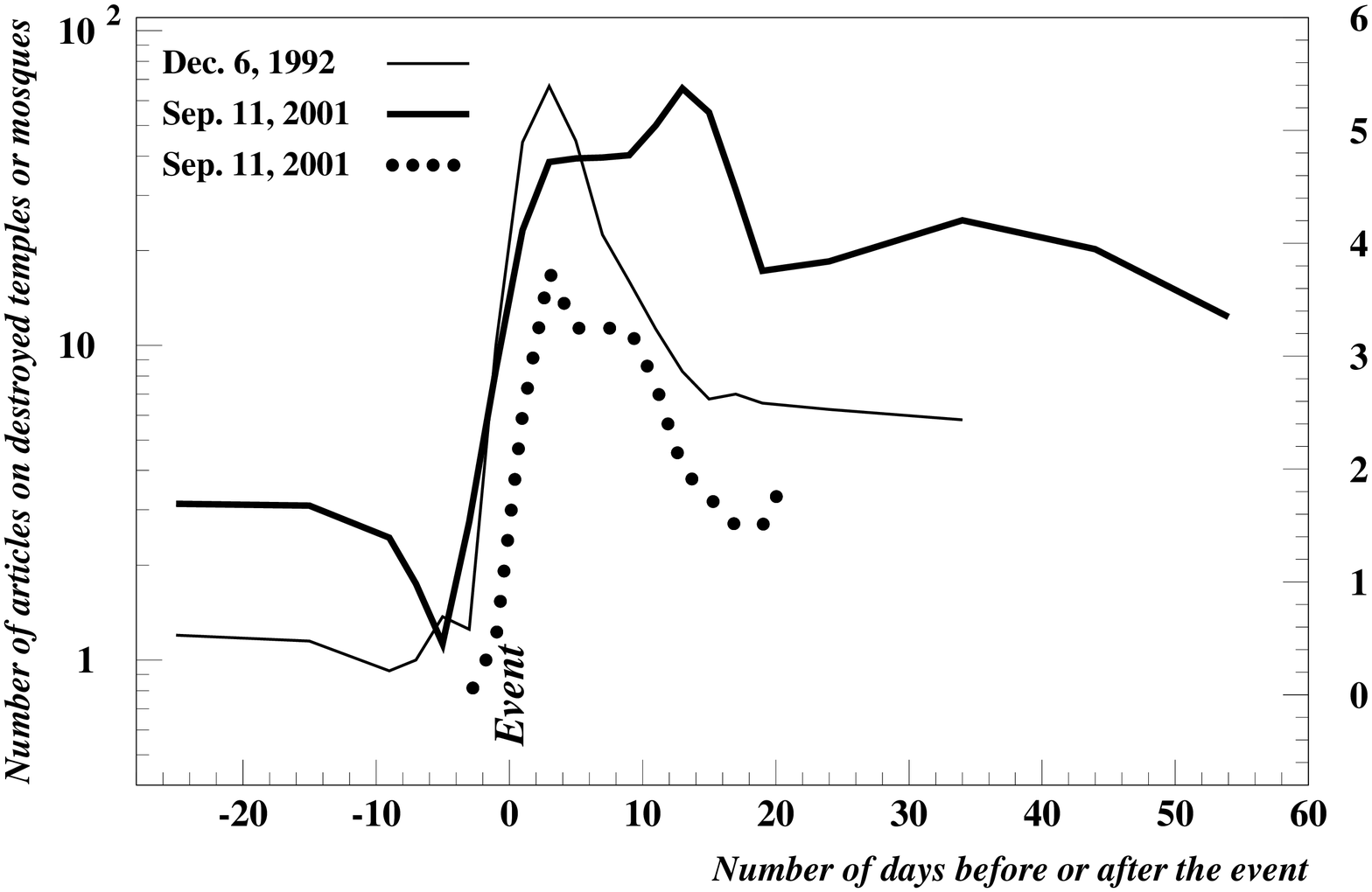}}
     {\bf Fig.1: Aftershocks of two critical events.}
{\small December 6, 1992 was marked by the destruction of
the Ayodhya mosque in India which sparked a wave of anti-Hindu
reactions; September 11, 2001 was marked by the destruction
of the Word Trade Center in New York which sparked a wave
of anti-Islamic reactions. The origin of the
horizontal scale corresponds to the day when the critical
event occurred. The two solid lines
show the number of articles writing on the destruction of
Hindu temples or mosques respectively (scale on the left-hand side);
the dotted line shows the number of mosques
actually destroyed or damaged (scale on the right-hand side).}
  \end{figure}
%% --------------------------------------------------

Several observations can be made. (i) There is a noise background
(similar to the cosmic rays background in particle detectors)
especially for the curve about mosques
which means that mosques are being destroyed almost at any
time; this particularly happens in Nigeria where there is an
almost permanent confrontation between Muslims in the north
and non-Muslims in the south. (ii) A first maximum is reached
about 3 days after the event. Thus, there were 102 articles in the
interval 8-9 December 1992 and 54 articles in the
interval 13-14 September 2001; however, in the case of September 11
there is a second peak about 11 days after the event.
\qpar

Figure 2 provides a second test of the significance of the
number of newspaper articles as an estimate of the
anti-Arab tension. It compares the number of articles
(albeit over a longer time interval than in Fig.1)
with the monthly number of anti-Arab aggressions in
California. Although the two curves
represent fairly different things, they are fairly
parallel at least in the two and three month
range covered by this figure; this suggests that (i) California is a good
proxy for the Western world and (ii) the number of articles
is a good proxy for the actual number of aggressions
in confirmation with our previous observation.

%%-----------------------------------------------
%%%% Fig.2
   \begin{figure}[htb]
    % \centerline{\psfig{width=17cm,figure=roehnersa2.eps}}
     \centerline{\psfig{width=17cm,figure=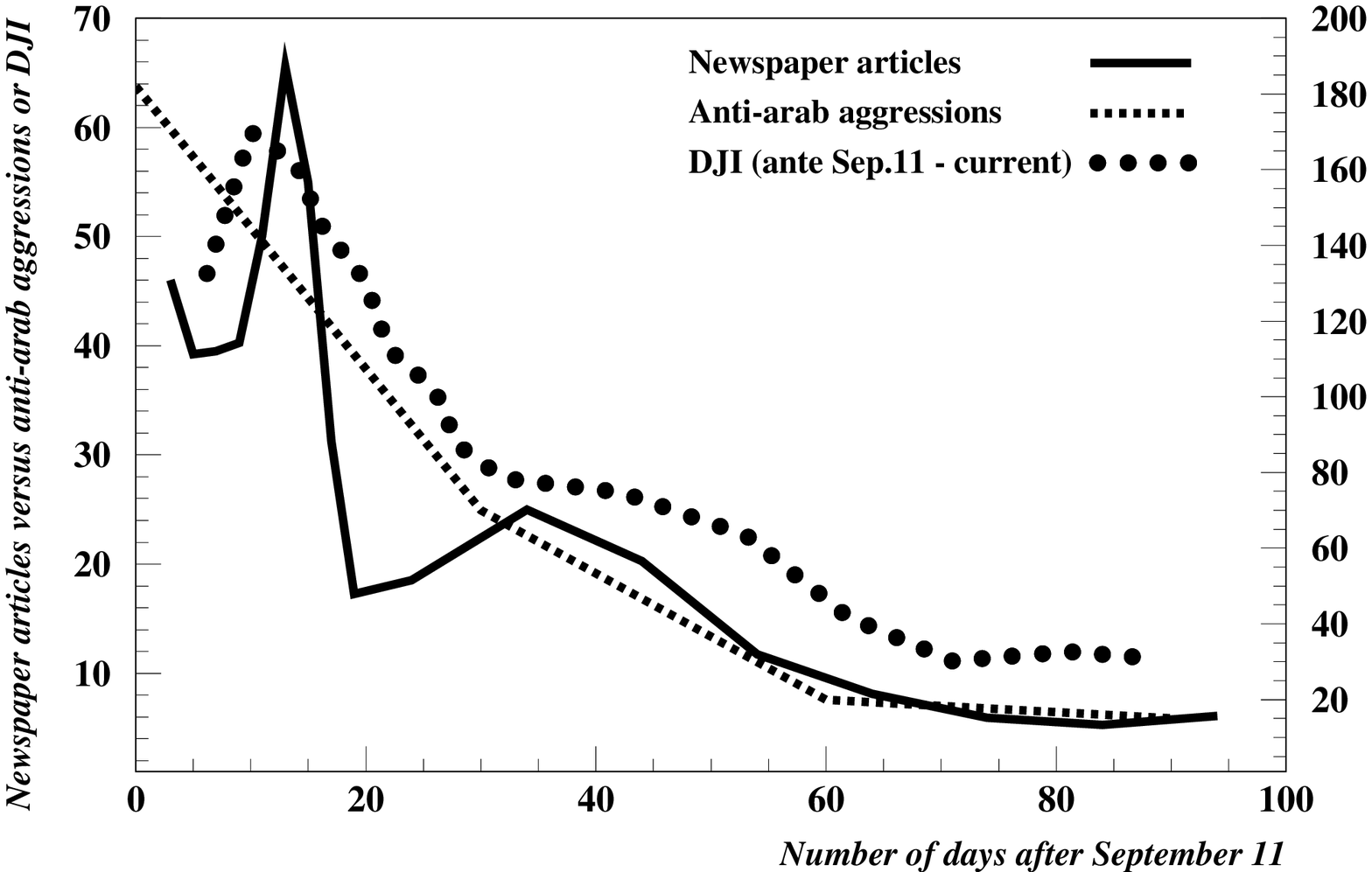}}
     {\bf Fig.2: Relaxation curves after the shock of
September 11.}
{\small The solid line curve is the same as in Fig.1 but
over a larger time interval; the broken line (scale on the
right-hand side) shows the number of anti-arab aggressions
in California in the three months after September 11;
the dotted line shows the changes in the level of the
Dow Jones Index with respect to its pre-Sep.11 level as given by
the difference DJI(pre-9/11)-DJI(current).
The tails of all three curves are well-approximated by
power laws $\sim 1/t^{\alpha}$, with exponents $\alpha $ comprised between
-1.4 and -2.2: $ \alpha _1=-1.8 \pm 0.7$ (newspaper articles),
$\alpha_2=-1.4 \pm 0.5$ (anti-arab agressions)
and $ \alpha _3=-2.2\pm 1.6 $ (DJI).}
\qL
{\it Source: California's Attorney General Office,
published in the
San Jose Mercury News, 11 March 2002.}

  \end{figure}
%% --------------------------------------------------

\qpar

In a second stage, we read the articles or at least
those which, according to their title, seemed particularly
relevant. The objective was to count the actual number of
buildings (temples or mosques) destroyed or damaged and
where they were located. One would
expect that the number of articles devoted to a specific
arson attack against a mosque depends upon the attention
newspaper give to this kind of events. This second
investigation allowed us to see whether the relationship
between the number of articles and the number of damaged
buildings remains constant in the course of time. The results
are represented by the dotted curve in Fig.1 and by the data
in Table 1a,b.

%%%%%%%%%%%%%%%%%%%%%%%%%%%%%%%

% TABLE 1ab

\begin{table}[htb]

  \small

\centerline{\bf Table 1\ Number of temples/mosques destroyed}

\vskip 3mm
\hrule
\vskip 0.5mm
\hrule
\vskip 2mm

\vskip 3mm
\centerline{\bf Table 1a\ Number of Hindu temples destroyed after
December 6, 1992}
$$ \matrix{
\tvi
\hbox{Country}\hfill & \hbox{Number destroyed} & \hbox{Some locations}
\hfill \cr
\noalign{\hrule}
\qth
\hbox{Afghanistan}\hfill & \geq 3  &  \cr
\hbox{Bangladesh}  \hfill & \geq 30  &  \cr
\hbox{Britain}\hfill & \sim 15 & \hbox{\small Birmingham, Blacklake,
   Bolton,} \hfill \cr
\hbox{}\hfill &  &  \hbox{\small Coventry, West Bromwich}\hfill \cr
\hbox{Canada}\hfill & \phantom{>} 1  & \hbox{Montreal} \hfill \cr
\qtb
\hbox{Pakistan}  \hfill & \geq 50 & \hbox{Karachi, Lahore, Multan,
Rawalpindi} \cr
\noalign{\hrule}
} $$
\vskip 1.5mm
Note: For Afghanistan, Bangladesh and Pakistan the source provides
only partial coverage.
\qL
Source: Lexis-Nexis (newspaper database)
\vskip 6mm

\centerline{\bf Table 1b\ Number of mosques destroyed after September
   11, 2001}
$$ \matrix{
\tvi
\hbox{Country}\hfill & \hbox{Number destroyed} & \hbox{Some locations}
\hfill \cr
\noalign{\hrule}
\qth \hbox{Australia}  \hfill & 3 & \hbox{\small Brisbane (2), Gold
   Coast}
\hfill \cr
\hbox{Britain}  \hfill & 4 & \hbox{\small Bolton, Manchester, Oldham}
\hfill \cr
\hbox{Canada}\hfill & 6 & \hbox{\small Hamilton, Montreal, Vancouver}
\hfill \cr
\hbox{Netherlands}\hfill & 4  & \hbox{\small Nijmegen, Venlo, Zwolle}
\hfill \cr
\qtb
\hbox{United States}\hfill & 6 & \hbox{\small Austin, Chicago,
   Cleveland, San Diego} \hfill \cr
\noalign{\hrule}
} $$
\vskip 1.5mm
Notes: With respect to Islamic population in each country,
the number of destroyed mosques per million are as follows
(estimated Islamic population in millions is given within parenthesis):
Australia (0.28) 10.; Britain: (1.5) 2.7; Canada: (0.58) 10.;
Netherlands (0.92) 4.3; United States (6) 1.0.
\qL
Source: Lexis-Nexis (newspaper database)
\vskip 2mm

\hrule
\vskip 0.5mm
\hrule

\normalsize

\end{table}

%% --------------------------------------------------------------

Let us discuss these results in more detail.
\qL
The dotted curve in Fig.1 shows the number of mosques which
were destroyed in the wake of September 11. It is interesting
to notice that its overall shape is fairly parallel
to the one for the number of articles; in particular the bulk
of the attacks and of the articles occurs between 3 and
15 days after September 11.

\qL
As can be seen in Table 1a, most of the attacks against Hindu
temples occurred in Bangladesh and Pakistan which is of
course not surprising since these countries are close to
the place of occurrence of the critical event both culturally
and geographically. However, the newspapers contained in
Lexis-Nexis do not provide accurate information about these
attacks. One account reads: ``At least 50 Hindu places of
worship were demolished throughout Pakistan during the last
two days'' (Guardian, December 9, 1992); neither the exact number nor
the locations are given. One would need accounts from local
Pakistani journals but these journals either are not
available on Lexis-Nexis or they are not in English.
This is the reason why we did not include a curve for
attacks against Hindu temples in Fig.1.
The situation is much more favorable for the destruction
of mosques for in this case most of them occurred in
western countries and are documented in detail. Not only do
we know the number and location with accuracy but we
are also told about the circumstances of the attack (petrol
bomb, explosive, car driven through the main gate, etc.).
Naturally, we cannot be sure that some attacks were not left
unrecorded especially those which did not bring about
great damages.
\qpar

{\bf Identification of cracks}. Obviously we may expect
destruction of Hindu temples only in those places where
Muslim and Hindu communities live side by side. The data
in table 1a show that there is a marked Hindu/Muslim divide in
Bangladesh and Pakistan and it also suggests
that the divide is more pronounced in Britain than in Canada.
This seems consistent with what we
know about communal violence in the two countries:
in Britain there were many serious riots
whereas in Canada there were only a few outbreaks.

\qL
Surprisingly, after September 11, there have been
more attacks in Canada than in Britain. Is this the
effect of geographical proximity or of burgeoning
communal cracks in Canada? At this point, we leave the
question open. With respect to its size, the Netherlands
has seen a fairly large number of attacks against mosques.
This can be put in relation with the fact that such
attacks have been fairly common in previous years. For
instance, during the first four months of 1992, there have
been 10 attacks (Donselaar 1993), an observation which
confirms that critical events are indeed able to reveal
pre-existing cracks. Private communications to the authors
from the Ministry of Justice of the Netherlands confirm
that there are deep concerns about the integrity of
the social tissue in the Netherlands, a fact illustrated
more recently
on the political scene by the rapid rise and then
assassination of the rightist politician Fortuyn in May 2002.

\qL
It may also be of interest to observe that after
September 11, there were no attacks in France, Germany or
Sweden, in spite of the fact that these countries have
substantial islamic minorities.
\qpar

{\bf Robustness} In order to see whether the critical
event approach is indeed able to reveal {\it  structural}
crashes, it is important to test the robustness of the
previous observations. There is an event which allows a further
test.
In November 1990, thousands of Hindu extremists made a
first attempt to storm the Ayodhya mosque, but they were
stopped by security forces (24 people were killed).
This failed attempt nevertheless triggered anti-Hindu
reactions which were similar to those after December, 6,
1992, albeit of a smaller scale. For instance, in
Pakistan, at least 4 Hindu temples were set on fire, and
in Bangladesh at least 300 Hindu homes and shops
were torched (Japan Economics Newswire, 4 November 1990).

\qI{The method of critical events in economics \label{criticalecon}}

In the previous sections, we have investigated the response
of social systems to (more or less) ``controlled'' shocks;
it is natural to ask if the same methodology also
applies to economic systems?
\qL

A first example can be found in Fig.2. It shows
that the decrease and  subsequent increase
of the Dow Jones Index (the curve  on Fig.2 is inverted
because it represents the difference between the level
prior to September 11 and the current level) fairly
closely follows the social jolt provoked by the attacks on
the World Trade Center. This observation comes as a
confirmation of similar ones made in previous
publications (Roehner and Sornette, 2000) which also showed
the strong impact that social manias have on stock prices.
\qL

Yet, economics is primarily concerned with medium-
and long-term phenomena, whereas the critical event
approach seems to focus on short-term responses.
In order to show that the notion can be extended
to multi-annual phenomena, we propose the following
example.
\qpar

In times of double-digit inflation rates, investors try to set up
hedging strategies by investing in diamonds,
precious metals, collectibles and other tangible assets
which are thought to provide safe shelters.
For instance, there was a price peak in antiquarian books in the
United States in the 1970s which matched the high inflation
rates (19 percent in 1974);
similarly in the U.K., the stagflation years
(with an inflation rate of 24 percent in 1974) induced a
price peak for postage stamps;
%there was also a price peak for
%postage stamps in France in the early 1940s which was a consequence
%of the high inflation rate (46 percent in 1945).
In short, such
episodes give us the opportunity to study how the markets of
tangible assets react to a common shock. Fig.3a permits to
estimate the response function of diamonds
and several precious metals. It should be noted that the prices of
other metals such
as copper or aluminum reacted only very moderately to the same
shock. The Ayodhya and September 11 episodes permitted to identify
social fractures; in much the same way, the double-digit inflation
episodes allow us to probe the extent to which a given commodity
is considered a safe shelter.

%%%% Fig.3a,b
   \begin{figure}[htb]
   %  \centerline{\psfig{width=17cm,figure=roehnersa3.eps}}
    \centerline{\psfig{width=17cm,figure=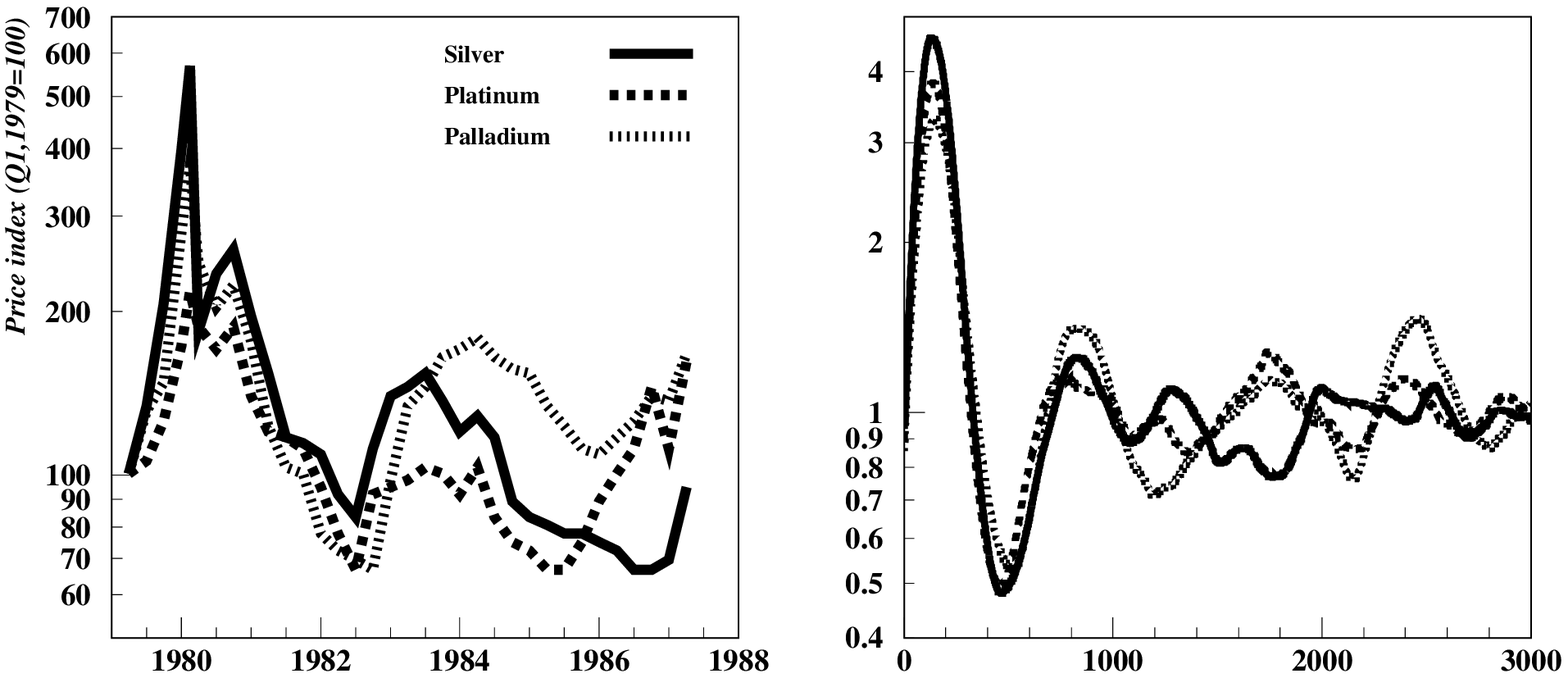}}
     {\bf Fig.3a: How the price of precious metals react
to high inflation.}
{\small The price series for diamonds, gold or cobalt are
very similar; they have been omitted for the sake
of clarity; in contrast the price of non-precious metals such as
copper or aluminum did not experience a peak.
It can
be observed that the strong correlation between prices
ends after the burst of the speculative bubble; in normal times, the
correlation between the prices of precious metals is very weak.
{\it Sources: The Economist (25 April 1987, Chalmin (1999).}}
\qL
{\bf Fig.3b: Three realizations of a shock-driven second order autogressive
process.}
{\small In addition to the standard noise term, the equation
contains a delta function $ \delta _{t0} $ which triggers the
peak at time $ t=0 $; the equation reads:
$ a_0X_t+a_1X_{t-1}+a_2X_{t-2}=\delta _{t0} + \epsilon _t $; the
parameters $ a_0, a_1, a_2 $ were chosen so that the relaxation time
is of the order of the width of the price peak, namely:
$ a_0=3434.15,\ a_1=-6847.58,\ a_2=3413.79$. Note that $a_0 \approx a_2$
and $a_1 \approx a_0+a_2$, which corresponds to a discrete
wave equation with weak damping.
As expected the realizations become independent stochastic
processes as soon as the effects of the initial shock vanish.
The main purpose of this simulation is to show that
no permanent exogenous
factor is required in order to generate such price peaks (an initial
shock is enough).
}

  \end{figure}

\qL
It is interesting to observe that once the speculative bubble
is over the collective phenomena ends and the price series progressively
resume their behavior as (more or less) independent random variables.
\qL

The whole
episode can fairly well be modelled through an autoregressive
process triggered by a discrete delta distribution as shown in
Fig.3b. This model confirms our interpretation, namely that there
is an ongoing competition process between the common response
triggered by the delta function ($ \delta _{t0} $)
shock on the one hand, and the
random fluctuations induced by the noise term $ \epsilon _t $.

\qI{Extension of this approach to pre-Internet times \label{criticalhisto}}

An important message of this paper is that the Internet
opens a new era for quasi-experimental research in the social
sciences. In the previous sections, we have explained
why this progress may be of pivotal importance and we
have shown on actual examples how the Internet allows
us to track the worldwide effects of critical events in
a fairly systematic way. However, it would be unfortunate
if this approach could be used only over the last 20 years.
In this section we show that, thanks to appropriate sources
which provide broad coverage, it can at least in some cases
be extended to the pre-Internet era as well.
\qpar

Hundred years ago at a time when the British Empire
extended over Australia, Canada, parts of Africa, India and
Ireland, the {\it Times} of London provided detailed information
about all these areas. Each of its issues comprised as many
as 20 or 30 pages which was quite exceptional at that
time (most other European newspapers had less than 10 pages).
In short, due to its broad coverage the {\it Times} could be seen
as a kind of precursor of the Internet; incidentally, it also published
many letters to the editor which gave readers the
opportunity of a feedback.
\qpar

How are we going to use
that observation device, in other words which critical
event should we select? No simple recipe can be given.
Usually one comes across a ``good'' set of events almost by chance.
The events
that will be used here occurred during the Boer War (1899-1902).
The first months of the war were marked  by a
succession of disasters for Britain. As more and more troops were
shipped to South Africa
the British army progressively
gained the upper hand. Two landmark events were the relief
of Ladysmith (28 February 1900) and Mafeking (18 May 1900), two
garrison towns
which had been besieged for several months. The news of
their relief was greeted in London with great excitement,
joyful celebrations and manifestations of patriotism. The
interesting question is how these news were received elsewhere
for instance in Glasgow, Belfast, Dublin, Montreal or Bombay.
Fortunately, the Times provides a detailed account for almost all
these places. The following excerpts give a feeling of the
range of reactions.

\qpar
\qci{
{\it London}\quad A large crowd had gathered in front of Mansion
house and in other places. The strains of the National Anthem,
``Rule Britannia'' and ``Soldiers of the Queen'' occasionally
raised above the tumult of thousand of voices. The balconies
were filled with people who moved not only flags, but also
blankets, table cloth, towels and various feminine garment. The
cab were blazing with Union Jacks. Groups of four or five men chartered
four-wheelers and mounted on top of them, they rode along the streets
waving flags. At night there were very general illuminations, and
there was a grand torchlight demonstration. The celebrations in
other English cities (e.g. Birmingham, Bristol, Manchester, Plymouth,
etc.) presented similar features.}
\qpar

\qci{
{\it Belfast}\quad The Nationalists [i.e. the Catholics] in many districts
were determined not to allow the night to pass without showing their
disloyalty to the Queen and their hatred for those participating
in the demonstrations. As the crowd were singing ``Bravo, Irish
Fusiliers'' they were met with a shower of stones thrown by a
mob of Nationalists. Fortunately the police drove the assailants
back and dispersed them.}
\qpar

Similar confrontations took place in Montreal and
Jersey (a French-speaking Channel island belonging to England).
For instance in the latter case, dirty water was thrown
on pro-British demonstrators
from windows in the French district; inflamed by the
pro-Boer sympathies of French residents,
the demonstrators wrecked the windows of several
hotels and restaurants. The reaction of the French districts in
Jersey was matched by similar reactions in France itself as
revealed by the following episode that occurred in
Brittany. On March 3, 1900, the steamer
Victoria entered the port of St Malo bearing all its flags. As it
came alongside, people gathered on the quay shouting ``Down with
the British, hurrah for the Boers.''
\qpar

Table 2a provides a broad summary of the reactions triggered by
the English victories. In London and other English
cities, they reveal a high level of national cohesion; in contrast
the reactions in Scotland, Ireland, Quebec or Jersey show
various degrees of a weaker cohesion.  In table 2, we used the
following cohesion scale:
\qL
$ \bullet $ Low cohesion (index=1): A substantial part
of the population expresses its disapprobation; confrontation
with loyalists.
\qL
$ \bullet $ Medium cohesion (index=2): Celebration by officials without
substantial popular
participation or opposition.
\qL
$ \bullet $ High cohesion (index=3): Celebration by officials
along with many enthusiastic people in the streets.
\qpar
As the question of national cohesion only serves for illustrative
purposes, we will content ourselves with this fairly crude scale.
However it should be noted that the evidence which is available in
newspapers would permit the construction of a more detailed index-scale.
\qpar

Table 2b reports similar observations for the celebration of
the Victory Day after World War I in the United States.
The disturbances between
the black and white segments of the population foreshadowed the
wave of riots which would erupt in 1919--1920. But at that time,
these warnings remained confined in the last pages of the newspapers
and did not attract the attention which they deserved.
Here is, we believe, an important lesson to learn from studying
the response function to critical events: the probe of the social
cohesion is an indicator and predictor of future evolution of the system,
and in some cases foreshadows the path to a critical transition.

%%%%%%%%%%%%%%%%%%%%%%%%%%%%%%%

% TABLE 2ab

\begin{table}[htb]

  \small

\centerline{\bf Table 2\quad Celebrations after victories}

\vskip 3mm
\hrule
\vskip 0.5mm
\hrule
\vskip 2mm

\vskip 3mm
\centerline{\bf Table 2a\ After the relief of
Ladysmith (28 Feb. 1900) and Mafeking (18 May 1900)}
$$ \matrix{
\tvi
\hbox{Place}\hfill & \hbox{Qualitative features} \hfill & \hbox{Index} \cr
\noalign{\hrule}
\qth \hbox{London}  \hfill & \hbox{Thousands of people waving flags
and singing patriotic songs are in the streets;} \hfill
& 3 \cr
\hbox{}  \hfill & \hbox{illumination of public
buildings and private houses, fireworks} \hfill & \cr
\hbox{Glasgow}\hfill & \hbox{Celebration by officials; rioting
shipyard employees invade Glasgow University}\hfill & 2 \cr
\hbox{Dublin}\hfill & \hbox{Streets covered with loyalists, nationalists
watch in angry silence}\hfill & 2 \cr
\hbox{Belfast}\hfill & \hbox{Confrontation between Catholics and
Protestants}\hfill & 1 \cr
\hbox{Jersey}\hfill & \hbox{French minority opposes celebrations; their
windows are smashed}\hfill & 1 \cr
\hbox{Montreal}\hfill & \hbox{Hoisting of the french flag above
the British flag on the office of the newspaper ``La Patrie''}\hfill & 1 \cr
\qtb \hbox{}\hfill & \hbox{it is torn down by loyalist students after
a confrontation with French-Canadian students}\hfill & 1 \cr
\noalign{\hrule}
} $$
\vskip 1.5mm
Notes: The description for London also applies to
other cities in England, as well as to Toronto, Cape Town or
Newfoundland. The index of social cohesion is defined in the text.
\qL
Sources: Times, Petit Journal, La Croix.
\vskip 6mm

\centerline{\bf Table 2b\ After the victory ending World War I (11 November
1918) in the United States}
$$ \matrix{
\tvi
\hbox{Place}\hfill & \hbox{Qualitative features} \hfill & \hbox{Index} \cr
\noalign{\hrule}
\qth \hbox{Downtown Manhattan (NY)}  \hfill & \hbox{Paper
and ticker tape tossed into the streets by wagonload;} \hfill
& 3 \cr
\hbox{}  \hfill & \hbox{Thousands of shipworkers stop work
for two days} \hfill & \cr
\hbox{Chicago (Ill.)}\hfill & \hbox{Riotous celebrations; seven
people killed}\hfill & 2 \cr
\hbox{Newport News (Va.)}\hfill & \hbox{Thousand of soldiers and sailors
wreck street cars and raid restaurants}\hfill & 1 \cr
\qtb \hbox{Harlem (in Manhattan) (NY)}\hfill & \hbox{Two thousand 
Negroes riot on Lenox Av.}
\hfill & 1 \cr
\noalign{\hrule}
} $$
\vskip 1.5mm
Notes: The description for Manhattan also applies to
other cities of the North East or West. The
words in table 2b are taken verbatim from the historical headlines.
\qL
Source: New York Herald
\vskip 2mm

\hrule
\vskip 0.5mm
\hrule

\normalsize

\end{table}

%% --------------------------------------------------------------

This example shows that by focusing on a specific kind of events,
in this case victory celebrations, one is in a position to probe the
loyalism of various parts of a country in a comparative way.
The last two words are of cardinal importance; let us briefly
explain why. Needless to say, the feelings of Scottish, Irish, or
French Canadian people also manifested themselves through
various separatist disturbances (demonstrations, riots, strikes, and so
on) but in these cases one faces isolated events which cannot be
compared in any meaningful way. For instance, the Easter Rising of 1917
in Dublin is not matched by similar events in Glasgow or Montreal.
In contrast, the critical event approach permits to compare the
responses of several segments of the population to the same event.

\qpar

Is it possible to extend this approach to even earlier periods
of time? Once again, the answer to this question is conditioned by
the existence of sources which are able to provide sufficiently
broad coverage. The {\it Times} was founded in 1785 but at that time
and  until the mid-nineteenth century it had only a few pages
and is therefore unable to provide the coverage we are looking for. However,
we should not give up completely.
Indeed for earlier times, there is a favorable circumstance which
comes to our help.
As one knows, back in the 18th century, French was widely
spoken and used by the aristocracy throughout Europe. It played the
role of a universal
language of communication that English plays
nowadays. As a result, newspapers giving local news in French were
published in many European cities. Table 3 lists some of these
papers. This opens, at least in principle, the possibility
of doing for the 18th century the kind of investigation
described in previous sections.

%%%%%%%%%%%%%%%%%%%%%%%%%%%%%%%

% TABLE 3

\begin{table}[htb]

  \small

\centerline{\bf Table 3\quad Newspapers published in French language in
European cities}

\vskip 3mm
\hrule
\vskip 0.5mm
\hrule
\vskip 2mm

$$ \matrix{
\tvi
  & \hbox{Time period} \cr
\noalign{\hrule}
\qth \hbox{Gazette de France (Paris)}  \hfill & 1631-1792 \cr
\hbox{Gazette de Leyde} \hfill & 1677-1811 \cr
\hbox{Gazette d'Amsterdam} \hfill & 1688-1795 \cr
\hbox{Gazette de Berne} \hfill & 1689-1787 \cr
\hbox{Gazette d'Utrecht} \hfill & 1689-1787 \cr
\hbox{Gazette de Cologne} \hfill & 1734-1757 \cr
\hbox{Gazette de Vienne} \hfill & 1757-1792 \cr
  & & \cr
\qtb \hbox{Common time period (except last two)} \hfill & 1689-1787 \cr
\noalign{\hrule}
} $$
\vskip 1.5mm
Notes: Back in the 18th century
``Gazette'' was the common French word for newspapers. Note that
the ``Gazette d'Amsterdam'' has been edited in electronic format
(CDROM) in the late 1990s; this makes it much more accessible;
others should become available electronically in the near future.
\qL
Source: Scard (1991)
\vskip 2mm

\hrule
\vskip 0.5mm
\hrule

\normalsize

\end{table}

%% --------------------------------------------------------------

\qI{Tests of the impact of an exogenous shock in the Sznajd model of 
consensus build-up
\label{secstau}}

Correlations do not mean cause and effect. If event A happens directly after
event B, it is not yet proven that B was caused by A. As D. Stauffer
humoristically points out ``The birth rate and the
number of storks in Germany both diminished strongly in recent 
decades, ``proving''
that babies are brough by the stork.'' The events of Sept. 11, 2001, or
Ayodhya 1992 seem not only correlated to but also the cause of the described
reactions. But was the ``Russian crisis'' of August 1998 on the stock markets
caused by president Yeltsin's announcement that Russia cannot service 
its debts,
or was it caused by internal market forces merely triggered by Yeltsin?
Johansen and one of us have in the past suggested the second scenario
(Johansen and Sornette, 1999).
A proper study thus needs
fully reproducible phenomena, as often done in physics as well as in computer
simulations. One first studies the system without one special event, and then
repeats the experiment (or simulation) including the special event. The
difference in the two simulations is the effect caused by the special event.
For continuous variables, this is chaos theory with Lyaponov 
exponents; are they
discrete (and thus neither linear nor nonlinear), it is called damage spreading
by physicists since 1986, but invented by S. Kaufman 1969 for genetics
(influence of a single mutation). In this spirit,
we propose to cast some more light on the impact of an endogenous shock
in a social system by using a simple model of consensus formation in a
population of interacting agents. We borrow from the ``consensus'' literature
which asks when and how
a complete consensus may emerge from initially diverging opinions.
Of course, the use of a model of consensus formation can only capture
a (probably small) part of the complex social interactions
at work in the different examples discussed above.
But, as previous applications of statistical physics
have shown (see Stauffer (2003) and references therein), even simple
models may go a long way in helping select the important variables
and parameters by a slow and thorough process of
selection/mutation/death of the competing models.

Here, we use perhaps the simplest model of consensus build up, the
Sznajd model (Sznajd-Weron and Weron, 2003)
in which a pair of neighbouring agents on a square
lattice convinces its six neighbours of the pair opinion if and only if
the two agents of the pair share the same opinion.
This basic version of the Sznajd model
with random sequential updating always leads to a consensus,
even if more than two opinions are allowed or for higher dimensions. Let us
call $p$ the initial fraction of opinions equal to $+1$. Then, the
consensus is $+1$ if $p > 1/2$ and $-1$ if $p<1/2$.
The time needed to reach a complete consensus fluctuates widely (see below).

Since the case $p<1/2$ is symmetric to the case $p>1/2$, it is sufficient
to consider only the latter. In our simulations, we consider a two-dimensional
lattice of $N=50 \times 50$ agents with periodic boundary conditions.
In absence of a perturbation, a given system will evolve with probability $1$
to the consensus ``magnetization'' $M=+1$. Let us consider such a system
and simulate its time evolution. Let us clone it and then apply a shock at
some time $t_s=10$, say. The shock consists in
an instantaneous perturbation in the form of the sudden introduction
of a ghost-site as in Schulze (2003): at $t_s=10$, we introduce a ghost-agent
which is connected at random to the fraction $g$ of all sites; the 
ghost-agent has
the power to shift the opinion of the $N g$ agents to match her own opinion.
At the next time step $t_s +1$, the ghost-agent disappears and the
normal dynamics of the Sznajd model resumes.
The shock can alternatively be seen as impacting all $N$ agents
but with the ghost-agent being successful
in convincing only a fraction $g$ of them. Those agents
already of the same opinion as the ghost-agent are not modified.
This definition of a shock mimicks
a major disruption of the social network, which impact a significant
fraction of the total population, as in the real examples discussed above.

We test the two possibilities: the ghost-agent opinion is $+1$ or $-1$.
Since the initial
average initial opinion $(+1) \times p + (-1) \times (1-p) = 2p-1$ is
positive for our choice $p>1/2$, if the ghost-agent opinion is $+1$,
its action accelerates
the convergence to the consensus. If the ghost-agent opinion is $-1$,
this shock may delay reaching the same $+1$ consensus or reverse it.
To see this, let us calculate the fraction of $\pm 1$ opinions
after a negative shock (ghost-agent opinion $-1$). Let us call
$p_s$ the fraction of $+1$ opinions just before at the time $t_s$
of the shock. Since a negative shock only
change the opinion of positive opinions, the fraction of $+1$ opinions
after the shock is $p_s (1-g)$. The fraction of $-1$ opinions after the
shock is $(1-p_s) + g p_s$. The negative shock will be able to turn
around the consensus from $+1$ to $-1$ with certainty if
\begin{equation}
p_s (1-g) < (1-p_s) + g p_s
\end{equation}
and with probability $1/2$ (that is, in one out of two realizations at random)
if the inequality is changed into an equality. This gives the condition
on the shock strength
\begin{equation}
g > {2 p_s -1 \over 2 p_s}~, ~~~~~{\rm with}~~~p_s >1/2~,
\label{jjj}
\end{equation}
needed to change the consensus from $+1$ to $-1$.
This formula tells us that a tiny negative shock is enough to
flip down the consensus if it was fragile to start with ($p_s$ close 
to $1/2^+$),
while a shock as large as shifting one-half the opinion of the total
population is required in the case where the consensus $+1$ was almost reached.
We have checked the validity of this reasoning
and of expression (\ref{jjj}) with numerical simulations.

Our purpose now is to ask how the perturbed system may relax back following
its previous convergence to a consensus after the occurrence
of the shock. In this goal, we calculate
the difference $f_s^+(t)-f_0^+(t)$ between the fraction $f_s^+(t)$ of 
$+$-opinions
of the perturbed system and that
$f_0^+(t)$ of the unperturbed system and study its evolution as a 
function of time
for various values of the parameters. Note that
the difference $f_s^-(t)-f_0^-(t)$ between the fraction $f_s^-(t)$ of 
$-$-opinions
of the perturbed system and that
$f_0^-(t)$ of the unperturbed system is given by
\begin{equation}
f_s^-(t)-f_0^-(t) = - f_s^+(t)-f_0^+(t)~.
\end{equation}
In the cases considered here where the ghost-agent action has
a single sign and the dynamics of opinion changes is also always 
imitative, the number of agents whose opinions have been changed by the 
transient
action of the ghost-agent is the sum of the number of changes
from opinion $+$ to $-$ and of the number of changes from opinion $-$
to $+$:
\begin{equation}
D(t) = {|f_s^+(t)-f_0^+(t)| + |f_s^-(t)-f_0^-(t)| \over 2} = 
|f_s^+(t)-f_0^+(t)|~.
\label{jngkewl;}
\end{equation}
This number is usually called the damage variable. 
Our numerical simulations calculating directly
the number of sites affected confirm the exactness of relation (\ref{jngkewl;})
to obtain the damage variable. Of course, 
in the cases where simultaneous changes $+ \to -$ and $- \to +$ occur,
the damage variable is not given by (\ref{jngkewl;}) anymore: for instance,
if the whole $+$ population is changed into $-$ and the whole $-$ population
is changed into $+$, the damage variable is equal to $1$ while $D(t)$
defined by (\ref{jngkewl;}) can be identically zero (if the initial proportions
of $+$ and $-$ is $1/2$). 

The evolution of $D(t)$ allows us to measure the spreading of ``damage'' in the
network of acquaintance. The quantity $D(t)=|f_s^+(t)-f_0^+(t)|$ plays
the role of a response function.
In our simulations, we use the same
random sequential updating for the unperturbed and its perturbed clone.

Figure 4 shows the average over 10000 realizations of the
response function $D(t)=|f_s^+(t)-f_0^+(t)|$ (crosses)
for $p=0.55$ and a ghost-agent opinion $+1$ with impact $g=0.1$. The
response function $D(t)$ drops in a few time steps and then oscillates
over hundred of time steps in the range $1-2\%$. Figure 4 also shows
the fraction $P(t)$ (stars) among the 10000 realizations,
with non-zero $D(t)=|f_s^+(t)-f_0^+(t)|$ (``survivors''). 
$10000[1-P(t)]$ is thus the
number of realizations such that the unperturbed and perturbed systems
are undistinguishable at time $t$.
$P(t)$ is seen to decay approximately as a power law
from $t\approx 10$ to $500$ (the duration of our simulations). The average of
$D(t)=|f_s^+(t)-f_0^+(t)|$ conditioned on those realizations such that
$D(t)=|f_s^+(t)-f_0^+(t)|$ is still non-zero
is shown with the squares (it corresponds
to dividing the unconditional average of $D(t)$ (crosses) by $P(t)$ (stars)):
it first decreases, reaches a minimum and then increases!
This conditional average shows that the shock has two major effects. (i) First,
the average difference $D(t)=|f_s^+(t)-f_0^+(t)|$ decays as expected for
a response/relaxation function reacting to a Dirac perturbation. (ii)
For those realizations whose perturbation is still felt, the shock is
sufficiently large to modify so significantly the dynamics of the 
social network
towards consensus, that the difference between
the unperturbed and perturbed surviving realizations actually increases.
Of course, eventually it will drop to zero sharply as the last 
realization has its
unperturbed and its perturbed versions both reaching the same 
consensus, leading
to a zero difference.

\begin{figure}[htb]
%\centerline{\psfig{width=17cm,figure=roehnersa4.eps}}
\centerline{\psfig{width=17cm,figure=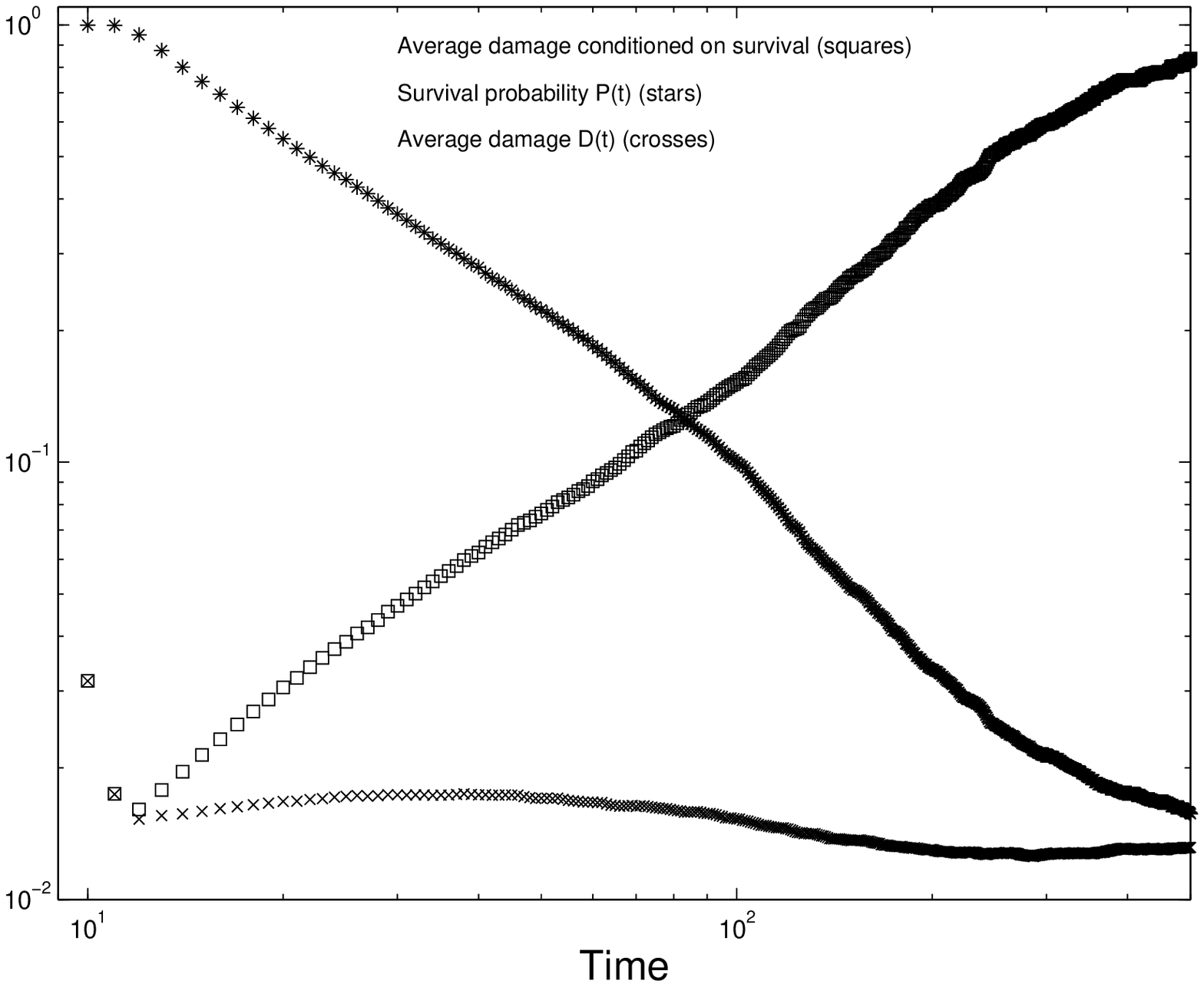}}
{\bf Fig. 4:}
{\small Average over 10000 realizations of the
response function or damage $D(t)=|f_s^+(t)-f_0^+(t)|$ (crosses)
for $p=0.55$ and a ghost-agent opinion $+1$ with impact $g=0.1$ (fraction
of agent that the ghost-agent convince of her own opinion).
$M_s(t)$ is
the average opinion (magnetization)  of the perturbed system and
$M_0(t)$ is the average opinion of the unperturbed system.
Stars ($*$): fraction $P(t)$ among the 10000 realizations which have
still different unperturbed and perturbed opinions
($D(t)=|f_s^+(t)-f_0^+(t)|$ is still non-zero)
at time $t$.
Squares:  average of $D(t)$ conditioned on those
realizations whose unperturbed and perturbed configurations are
different. See text for more details.
}
\end{figure}
\qL

Figure 5 is the same as Figure 4
for $p=0.65$ and a ghost-agent opinion $-1$ with impact $g=0.4$.
Comparing with Figure 4, here the negative shock is delaying
considerably the convergence
to the consensus but it is not strong enough to reverse it.

\begin{figure}[htb]
%\centerline{\psfig{width=17cm,figure=roehnersa5.eps}}
\centerline{\psfig{width=17cm,figure=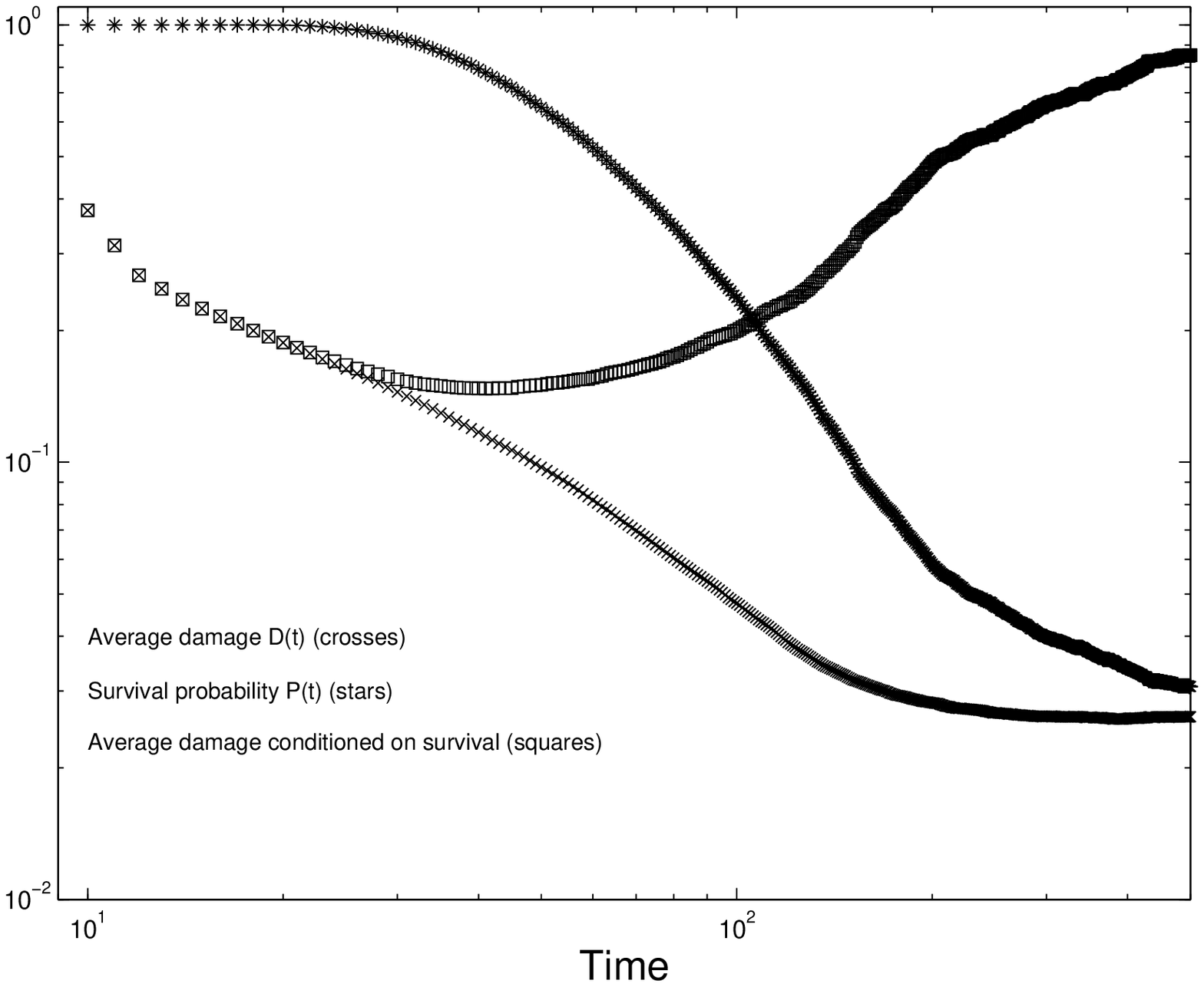}}
{\bf Fig. 5:}
{\small Same as Figure 4
for $p=0.65$ and a ghost-agent opinion $-1$ with impact $g=0.4$.
}
\end{figure}
\qL

Figure 6 shows the absolute value of the average over 10000 realizations of the
response function $D(t)=|f_s^+(t)-f_0^+(t)|$ (crosses)
for $p=0.65$ and a ghost-agent opinion $-1$ with impact $g=0.5$.
Comparing with Figures 4 and 5, the shock is now sufficiently strong
to reverse the consensus to $-1$ in the end. The damage $D(t)$
(crosses) no longer decreases but increases
with time. The fraction $P(t)$ (stars) among the 10000 realizations which have
still different unperturbed and perturbed opinions 
($D(t)=|f_s^+(t)-f_0^+(t)| \neq 0$)
is plateauing over a much longer time interval and then decays very 
slowly with time.
The absolute value of the average of $D(t)$ conditioned on those
realizations whose unperturbed and perturbed configurations are
different at time $t$ (squares) is now growing even more and converges to $+1$.
Figure 6 also shows the average of $|f_s^+(t)-f_0^+(t)|$ conditioned on those
realizations such that the difference of opinion
between the unperturbed and perturbed configurations goes to $-2: +1 \to -1$
at $t=500$ (empty circles):
these are those realizations whose consensus is changed
by the shock at or before $t=500$. The fact that they do not 
constitute the full sample
and remain below the unconditional average of 
$D(t)=|f_s^+(t)-f_0^+(t)|$ (crosses) is due
to the fact that a fraction of the 10000 realizations have still not joined.

\begin{figure}[htb]
%\centerline{\psfig{width=17cm,figure=roehnersa6.eps}}
\centerline{\psfig{width=17cm,figure=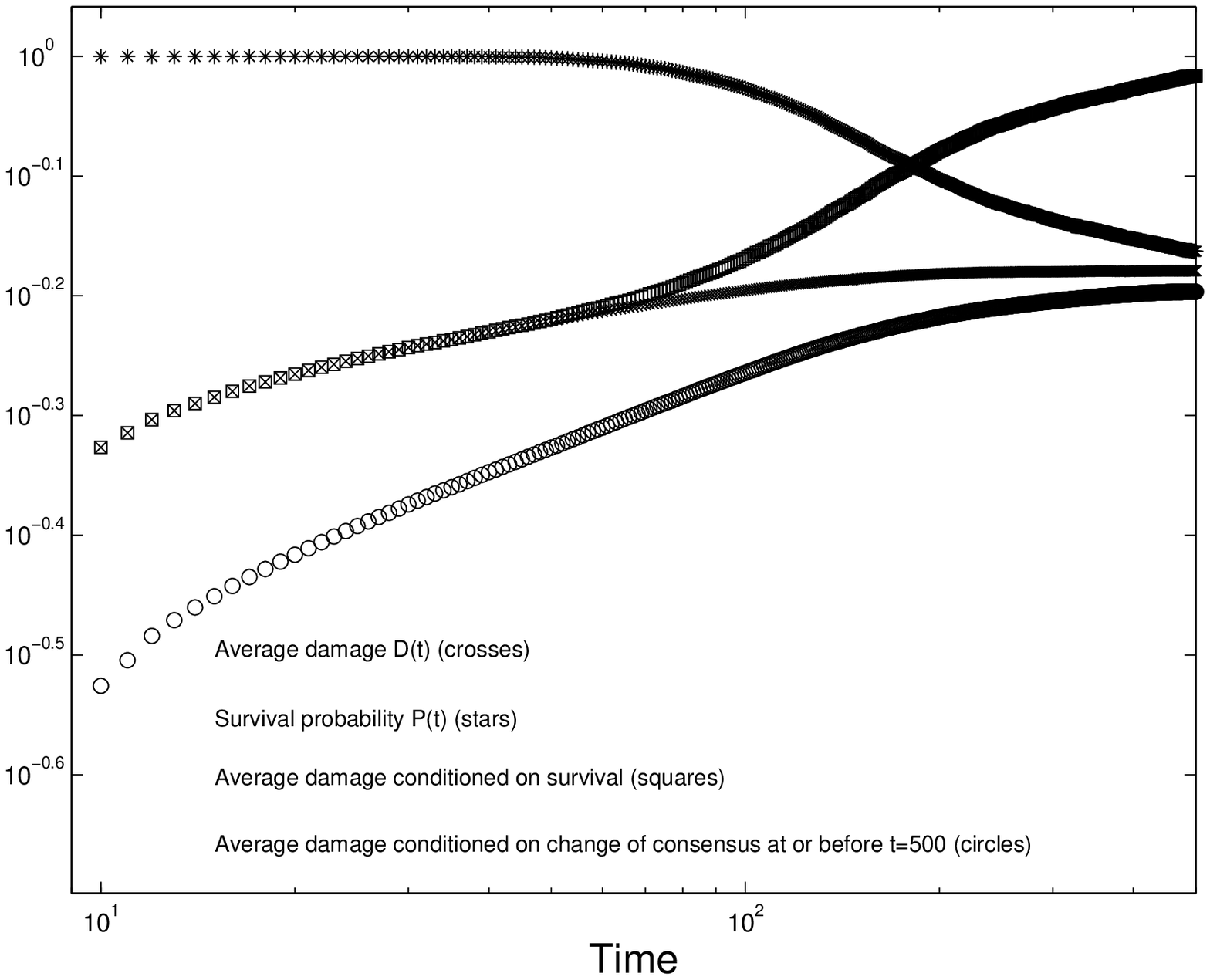}}
{\bf Fig. 6:}
{\small Same as Figure 4
for $p=0.65$ and a ghost-agent opinion $-1$ with impact $g=0.5$.
The empty circles are the average of the damage 
$D(t)=|f_s^+(t)-f_0^+(t)|$ conditioned on those
realizations such that the difference of opinion
between the unperturbed and perturbed configurations goes to $-2: +1 \to -1$
at or before $t=500$.
}
\end{figure}
\qL

How do these simulations compare with the empirical data
on the social response to acts of religious violence, to terrorism, 
to wars or to
economic shocks? First, one can observe a qualitative similarity between
the jump followed by the relaxation of the response function constructed
in the simulations and the measure of social reaction in the different
examples discussed above. However, in our empirical examples, we are not
able to really assess if a change of majority opinion has been triggered
by the shock, while the simulations demonstrate such a possibility if
the shock has a quality opposite to the pre-existing majority and is
sufficiently strong.
The simulations also emphasize the extremely large variability from
one realization to another, for the same control parameters. Since
social history provides us with only a limited number of examples,
each of them being unique in some characteristics, our simulations
suggest caution in over-interpreting some
characteristics of the social reaction which
could be the result of stochastic aspects of the response of the social system.
In this vein, future works will profit from the development of better
social models which could be calibrated to the data and used to 
assess what part of the observed
response function is universal and what part is idiosyncratic to local
pseudo-random occurrences.

\qI{Concluding remarks: structural versus temporal forecasts}

In the previous sections, we explained how the critical
event approach can help us unravel properties of a system
which remain hidden in normal times. How can
this new knowledge and understanding be tested?
The procedure could seem obvious: first derive some
forecasts, then confront them with the evidence. But one
should be aware that there are two kinds of forecasts. If
one predicts that quadrupling the length of a pendulum will
result in doubling its vibration period, this is what can
be called a structural prediction. Most of the predictions made
in physics are of that type. In contrast, the weather forecast
in the most common illustration of the second kind of prediction,
namely the temporal prediction.
In the social sciences, reliable temporal forecasts usually
are very difficult to make while structural predictions are
hindered by a lack of understanding of the fundamental mechanisms.
\qL

How does the above distinction apply to the examples that we set up
in the previous sections? Let us take an example.
Once one has identified the divide between
the French-speaking Canadian province of Quebec and the
English-speaking part of Canada, temporal prediction
would imply to forecast the date of a major
separatist disturbance. As one knows, such an episode took indeed place
but more than sixty years after the Ladysmith/Mafeking incidents.
The fact that it occurred in 1970 rather then twenty or fifty
years earlier (or later) has probably no simple explanation.
\qpar

In order to clarify the above distinction between structural
and temporal forecasts, it may be helpful to use a parallel with
forest fires. Although it is obvious that forest fires are
in some way related to climate dryness, dryness does not necessarily
lead to major forest fires. Apart from the dryness itself, other
factors must be present in order to bring about a major fire, such as
hot weather, strong wind, acts of negligence which may start
fires, and so on. As a result, large-scale forest fires will occur
infrequently, maybe only
every 20 or 50 years. To sum up, dryness will have a very low
correlation with forest fires especially if the fires are
observed over short periods of
time (say 20 years or less). Contrariwise, if one compares
climatic dryness and the amount of humidity contained in the
upper layer of the soil in different regions, one would
expect a high correlation. In short, by focusing on
structural effects, one will get fairly easily a clear
understanding, whereas when focusing on temporal forecasts
of major events, it will be difficult to come up with
clear-cut conclusions.
Unfortunately, in the social sciences there is
a tendency to concentrate on temporal events because
these are often the only ones for which any information is
available.
\qpar

Finally, we cannot close this paper without paying tribute to the
pioneering work of Stanley Milgram (1933-1984),
an American sociologist, whose
papers about
the lost letter technique or the small world experiment (Milgram 1970, 1977)
showed a promising direction of experimental research
which, unfortunately, did not have many continuators... until very
recently (see a repetition and enhancement of the small world 
experiment of Milgram
implemented on-line on the internet at
http://smallworld.columbia.edu/).

\qpar
{\bf Acknowledgements}: We are grateful to D. Stauffer for
suggesting the numerical simulations, for inspiration for the
introduction of section \ref{secstau} and for a careful reading
of the manuscript.

  \vfill \eject

\vskip 1.5cm

\centerline{\bf \Large References}

\vskip 1cm

\qparr
Dellago (C.) and Mukamel (S.) 2003:
Nonlinear Response of Classical Dynamical Systems to Short Pulses, Bull.
Korean Chem. Soc. 24 (8), 1107-1110.

\qparr
De Podesta (M.) 2002:
Understanding the properties of matter.
Taylor and Francis, New York.

\qparr
Donselaar (J. van) 1993: The extreme right and racist violence
in the Netherlands. in T. Bj\o rgo, R. Witte eds: Racist violence
in Europe. St. Martin's Press. New York.

\qparr
Durkheim (E.) 1897: Le suicide. F\'elix Alcan. Paris.

\qparr
Helmstetter (A.) Sornette (D.) and Grasso (J.-R.) 2003:
Mainshocks are Aftershocks of Conditional Foreshocks: How do foreshock
statistical properties emerge from aftershock laws,
J. Geophys. Res., 108 (B10), 2046, doi:10.1029/2002JB001991.

\qparr
Johansen (A.) and Sornette (D.) 1999:
Financial ``anti-bubbles'': log-periodicity in Gold and Nikkei collapses,
Int. J. Mod. Phys. C 10(4), 563-575.

\qparr
Johansen (A.) and Sornette (D.) 2004:
Endogenous versus Exogenous Crashes in Financial Markets,
in press in ``Contemporary Issues in International Finance,''
Nova Science Publishers
(http://arXiv.org/abs/cond-mat/0210509)

\qparr
Milgram (S.) 1970: The small world problem. in: J.W. McConnel, editor,
Readings in social psychology today.

\qparr
Milgram (S.) 1977: The individual in a social world: essays and
experiments. Addison-Wesley, Reading (Massachusetts).

\qparr
Morse (P.-M.) and Feshbach (H.) 1953:
Methods of theoretical physics (McGraw Hill, New York).

\qparr
Potter (S.M.) 2000:
Nonlinear Impulse Response Functions, Journal of Economic
Dynamics and Control 24 (10), 1425-1446.

\qparr
Putnam (R.D.) 2000: Bowling alone: The collapse and revival
of American community. Simon and Schuster. New York.

\qparr
Roehner (B.M.) 1999: The space-time pattern of price waves,
European Physical Journal B 8, 151-159.

\qparr
Roehner (B.M.) 2002a: Patterns of speculation. Cambridge University
Press. Cambridge.

\qparr
Roehner (B.M.), Syme (T.) 2002b: Pattern and repertoire in history.
Harvard University Press. Cambridge (Mass.)

\qparr
Roehner (B.M.), Rahilly (L.J.)  2002c: Separatism and integration.
Rowman and Littlefield. Lanham (Maryland).

\qparr
Roehner (B.M.), Sornette (D.) 2000: ``Thermometers'' of speculative
frenzy. European Physical Journal B 16, 729-739.

\qparr
Romer (D.) 1996: Advanced macroeconomics
(McGraw-Hill, New York, 1996.

\qparr
Scard (J.) 1991: Dictionnaire des journaux 1600-1789.
Presses Universitaires de Grenoble. Grenoble.

\qparr
Schulze (C.) 2003: Advertising effect in Sznajd marketing model,
Int. J. Mod. Phys. C 14, 95-98.

\qparr Schumpeter (J.A.) 1939:
Business Cycles: A Theoretical, Historical and Statistical Analysis of the
Capitalist Process (McGraw-Hill, New York).

\qparr
Sornette (D.) 2002: Why stock markets crash. Princeton
University Press. Princeton.

\qparr
Sornette (D.) and Helmstetter (A.) 2003:
Endogeneous Versus Exogeneous Shocks in Systems with Memory,
Physica A 318 (3-4), 577-591.

\qparr
Sornette (D.) Gilbert (T.) Helmstetter (A.) and Ageon (Y) 2004:
Endogenous Versus Exogenous Shocks in Complex Networks: an Empirical Test,
submitted to Physical Review Letters\\
(http://arXiv.org/abs/cond-mat/0310135)

\qparr
Sornette (D.), Malevergne (Y.) and Muzy (J.-F.) 2003:
What causes crashes?  Risk 16 (2), 67-71
(http://arXiv.org/abs/cond-mat/0204626).

\qparr
Stauffer (D.) 2003: How to convince others? Monte Carlo
simulations of the Sznajd model, American Institute of Physics Conference
Proceedings 690, 147-155.

\qparr
Sznajd-Weron (K.) and Weron (R.) 2003: Physica A 324, 437.

\qparr
Varshney (A.) 2002:
Ethnic Conflict and Civic Life: Hindus and Muslims in India
(Yale University Press).

\end{document}